\documentclass[a4paper,10pt]{article}
\usepackage[utf8]{inputenc}
\usepackage{authblk}

\usepackage{graphicx}
\usepackage{epstopdf}
\usepackage{todonotes}
\usepackage{lineno,hyperref}
\usepackage{upgreek}
\usepackage{units}
\usepackage{multirow}
\usepackage{array}
\usepackage{amssymb}	% Mathematische Symbole
\usepackage{amsmath}	% Mathematische Symbole
\usepackage{textcomp}
\usepackage{pdfpages}

\usepackage{caption}
\usepackage{subcaption}

\title{Mapping the depleted area of silicon diodes using a micro-focused X-ray beam}

\author[1]{Luise Poley}
\author[2]{Andrew Blue}
\author[3]{Ingo Bloch}
\author[2]{Craig Buttar}
\author[4]{Vitaliy Fadeyev}
\author[5]{Javier Fernandez-Tejero}
\author[5]{Celeste Fleta}
\author[6]{Johannes Hacker}
\author[7]{Carlos Lacasta Llacer}
\author[7]{Mercedes Mi\~{n}ano}
\author[3]{Martin Renzmann}
\author[3]{Edoardo Rossi}
\author[8]{Craig Sawyer}
\author[9]{Dennis Sperlich}
\author[3]{Martin Stegler}
\author[5]{Miguel Ull\'{a}n}
\author[10]{Yoshinobu Unno}

\affil[1]{Lawrence Berkeley National Laboratory, Cyclotron Road, Berkeley, USA}
\affil[2]{SUPA School of Physics and Astronomy, University of Glasgow, University Avenue, Glasgow, United Kingdom}
\affil[3]{Deutsches Elektronen-Synchrotron, Notkestra{\ss}e, Hamburg, Germany}
\affil[4]{Santa Cruz Institute of Particle Physics, University of California, High Street, Santa Cruz, United States of America}
\affil[5]{Centro Nacional de Microelectr\'{o}nica (IMB-CNM), Consejo Superior de Investigaciones Cient\'{i}ficas, Campus UAB-Bellaterra, Barcelona, Spain}
\affil[6]{Infineon Technologies Austria AG, Siemensstra\ss{}e, Villach, Austria}
\affil[7]{Instituto de F\'{\i}sica Corpuscular, CSIC-U. Valencia, c/ Catedr\'{a}tico Jos\'{e} Beltr\'{a}n, Paterna, Spain}
\affil[8]{Particle Physics Department, STFC Rutherford Appleton Laboratory, Harwell Science and Innovation Campus, Didcot, United Kingdom}
\affil[9]{Humboldt-Universit\"at zu Berlin, Newtonstra\ss{}e, Berlin, Germany}
\affil[10]{Institute of Particle and Nuclear Study, KEK, Oho, Tsukuba, Japan}

\begin{document}

\maketitle

\begin{abstract}

For the Phase-II Upgrade of the ATLAS detector at CERN, the current ATLAS Inner Detector will be replaced with the ATLAS Inner Tracker (ITk). The ITk will be an all-silicon detector, consisting of a pixel tracker and a strip tracker. Sensors for the ITk strip tracker are required to have a low leakage current up to bias voltages of \unit[-500]{V} to maintain a low noise and power dissipation. In order to minimise sensor leakage currents, particularly in the high-radiation environment inside the ATLAS detector, sensors are foreseen to be operated at low temperatures and to be manufactured from wafers with a high bulk resistivity of several $\text{k}\Omega\cdot\text{cm}$. 
Simulations showed the electric field inside sensors with high bulk resistivity to extend towards the sensor edge, which could lead to increased surface currents for narrow dicing edges.
In order to map the electric field inside biased silicon sensors with high bulk resistivity, three diodes from ATLAS silicon strip sensor prototype wafers were studied with a monochromatic, micro-focused X-ray beam at the Diamond Light Source (Didcot, UK). For all devices under investigation, the electric field inside the diode was mapped and its dependence on the applied bias voltage was studied.

\end{abstract}

\section{Introduction} 

For the Phase-II Upgrade of the ATLAS Detector~\cite{ATLAS}, its current Inner Detector will be replaced by the ATLAS Inner Tracker (ITk), which will consist of a pixel tracker and a strip tracker~\cite{ITk}. Silicon sensors for the future ATLAS strip tracker have been developed (\cite{ATLAS07}, \cite{ATLAS12}) to meet the challenging requirements of the sensor characteristics and stability in a high radiation environment. In order to achieve a low depletion voltage and thereby increase the readout signal, the wafer material used for silicon strip sensors is required to have a high bulk resistivity.

\begin{figure}
\centering
\includegraphics[width=0.8\linewidth]{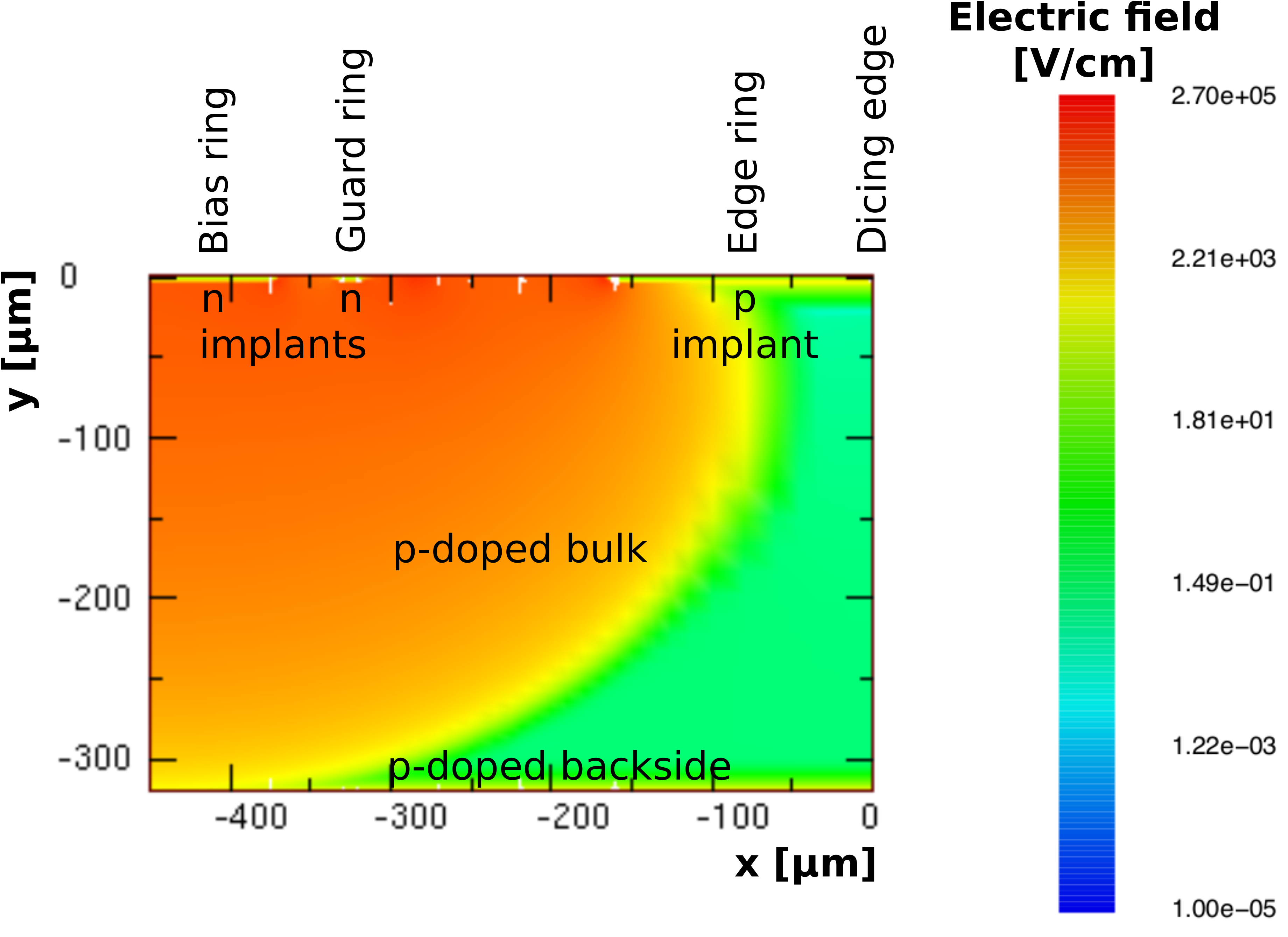}
\caption{Simulation of the electric field inside the edge region of a fully depleted silicon strip sensor with a bulk resistivity of {\unit[3]{k$\Omega\cdot$cm}} at a bias voltage of {\unit[-500]{V}}, simulated using ENEXSS~\cite{ENEXSS}. Positions of bias ring, guard ring, edge ring and dicing edge are indicated. The higher field strength areas of the depleted sensor volume is shown to extend towards the sensor dicing edge.}
\label{fig:sim}
\end{figure}

Simulations of the electric field inside a biased sensor have shown that the combination of the foreseen sensor parameters:
\begin{itemize}
 \item high resistivity material
 \item sensor doping profile
 \item dicing edges close to the active sensor area
 \item high bias voltage
\end{itemize}
could potentially lead to a breakdown in leakage current if it extends too far towards the dicing edge (see figure~\ref{fig:sim}). 
The simulation shows the edge region of a sensor (p-doped bulk) with a grounded bias ring (connected to an n-doped implant), guard ring (connected to an n-doped implant), edge ring (connected to a p-type implant) and a highly p-doped layer covering the sensor backside.
The dicing edge of a sensor is a highly degraded silicon surface with a sufficiently large concentration of free charge carriers to be electrically conductive. This conductive edge can lead to a short circuit between the sensor backside (to which high voltage is applied) and free charge carriers accumulating beneath the bias ring (grounded) and guard ring (floating), resulting in a significant increase of the leakage current. A highly p-doped edge ring is added to the sensor to prevent the formation of a conductive path between free charge carriers under the bias and guard ring and the dicing edge. The presence of a highly p-doped edge ring reduces the extension of the space charge region and the electric field towards the dicing edge.

Monitoring diodes were added to prototype wafer layouts which, despite different geometries and sizes compared to the full scale sensor, were designed to have edge regions representative of full size sensor edges. For this study, the electric field inside diodes from prototype wafers was studied by mapping their depleted areas.

\section{Diodes under investigation}

The diodes used for these measurements were included in wafer layouts for silicon strip sensors as test structures. Their wafer related characteristics (thickness of $\unit[310\pm10]{\upmu\text{m}}$, p-doping concentration in bulk and high resistivity bulk material) are the same as for full size ATLAS silicon strip sensors.

\begin{figure}
\centering
\begin{subfigure}{.28\textwidth}
  \centering
  \includegraphics[width=\linewidth]{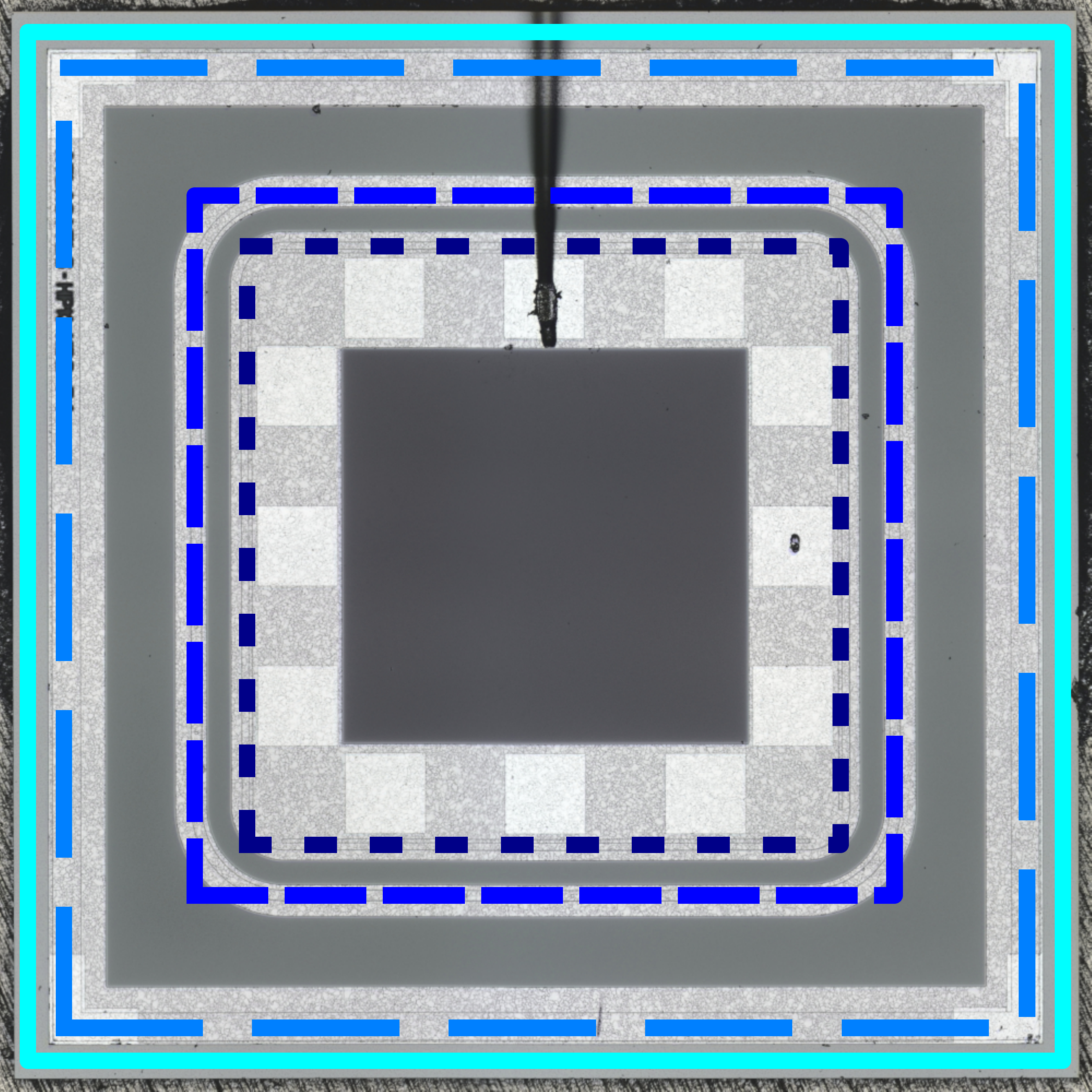}
  \caption{HPK diode with a size of {$\unit[2\times2]{\text{mm}^2}$} (HPK MD)}
  \label{fig:dioden_1}
\end{subfigure}
\begin{subfigure}{.28\textwidth}
  \centering
  \includegraphics[width=\linewidth]{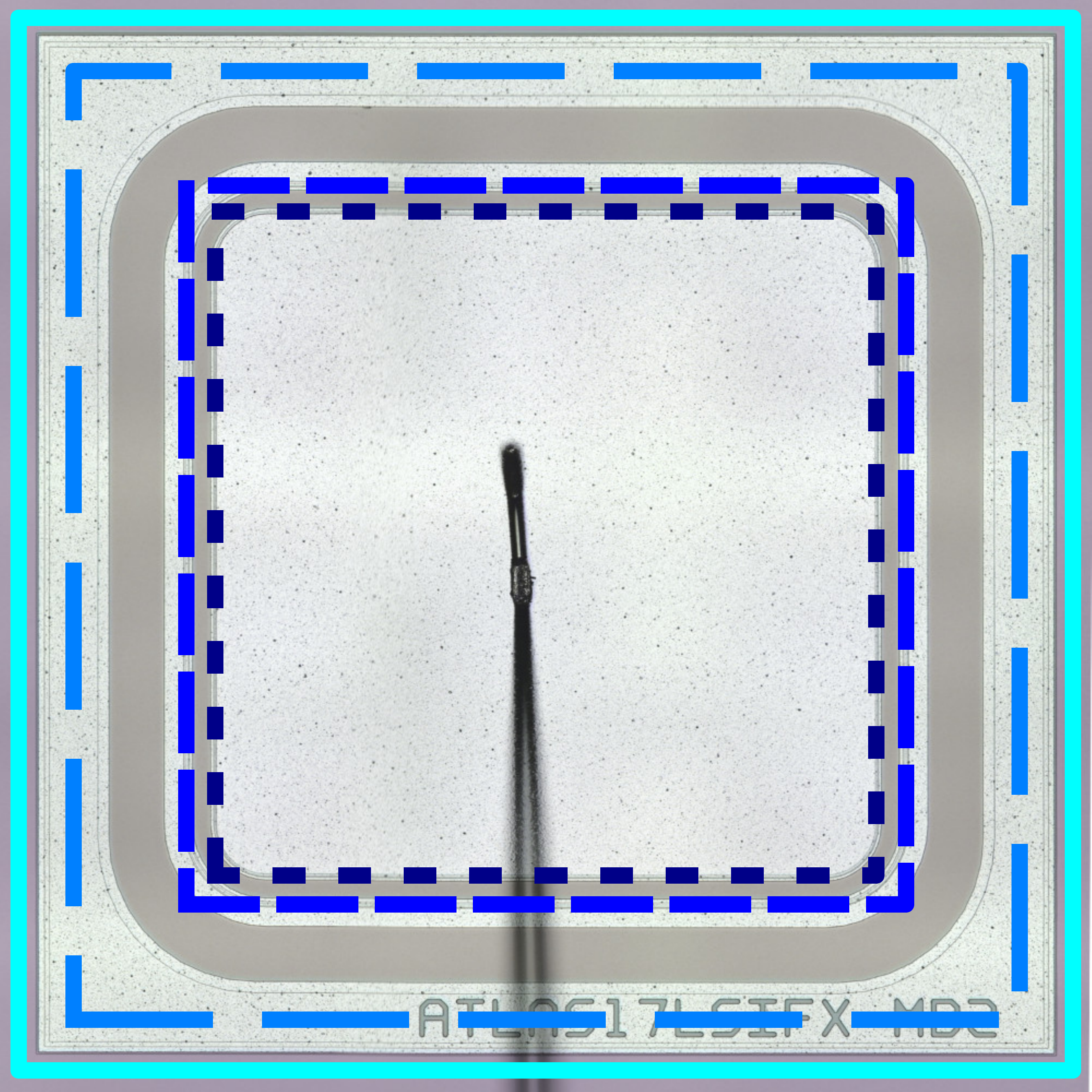}
  \caption{IFX diode with a size of {$\unit[2\times2]{\text{mm}^2}$} (IFX MD2)}
  \label{fig:dioden_2}
\end{subfigure}
\begin{subfigure}{.42\textwidth}
  \centering
  \includegraphics[width=\linewidth]{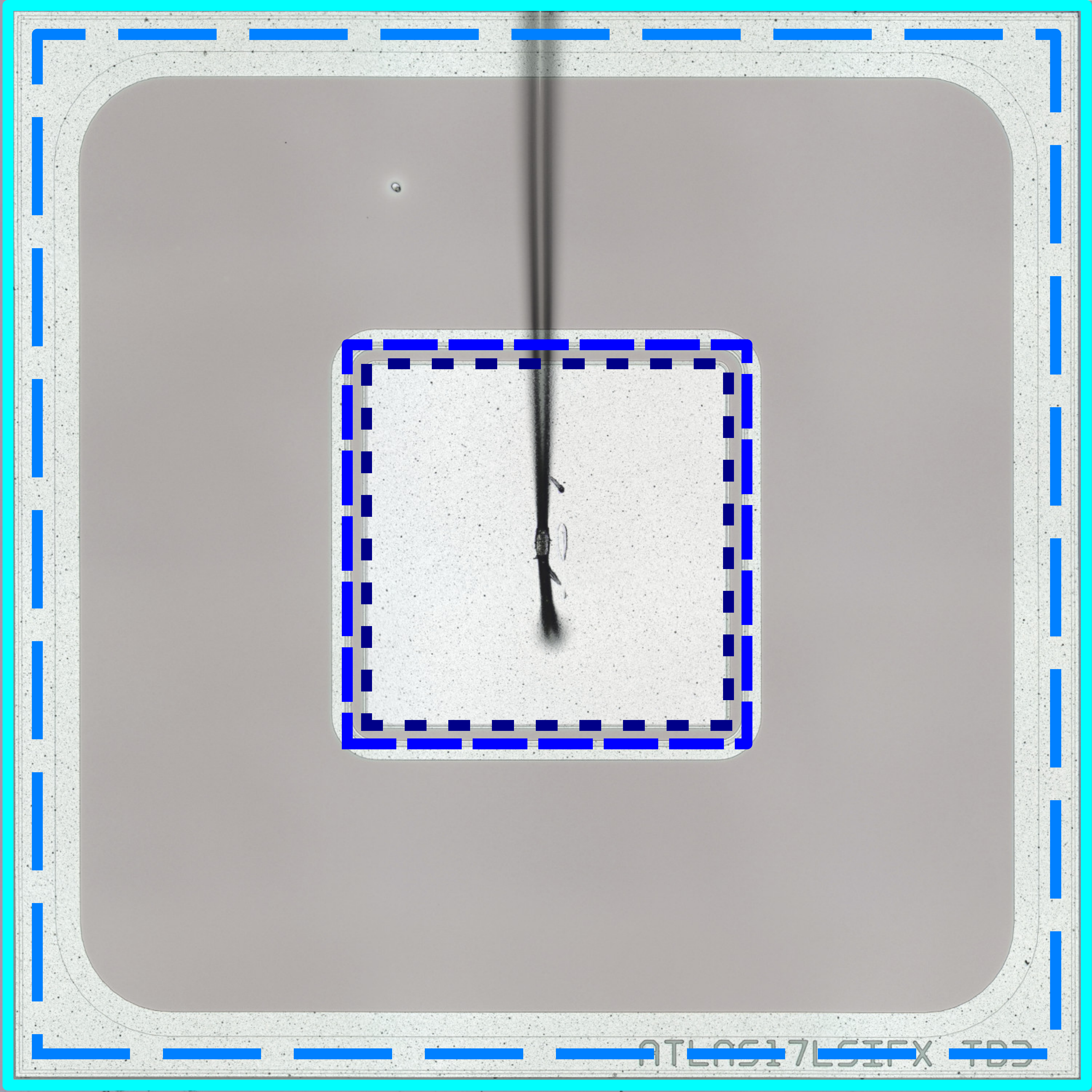}
  \caption{IFX diode with a size of {$\unit[3\times3]{\text{mm}^2}$} (IFX TD3)}
  \label{fig:dioden_3}
\end{subfigure}
\caption{Diodes used to study the electric field inside biased high resistivity material. Lines indicate approximate feature positions of each diode: edge of the diode implant (dotted dark blue line), guard ring (small dashed blue line), centre of the edge ring region (large dashed blue line) and nominal diode sizes (solid cyan lines, {$\unit[2\times2]{\text{mm}^2}$} or {$\unit[3\times3]{\text{mm}^2}$}). Wire bonds were attached to the diode surface in order to bias the diodes.}
\label{fig:test}
\end{figure}

\begin{figure}
\centering
\begin{subfigure}{.67\textwidth}
  \centering
  \includegraphics[width=\linewidth]{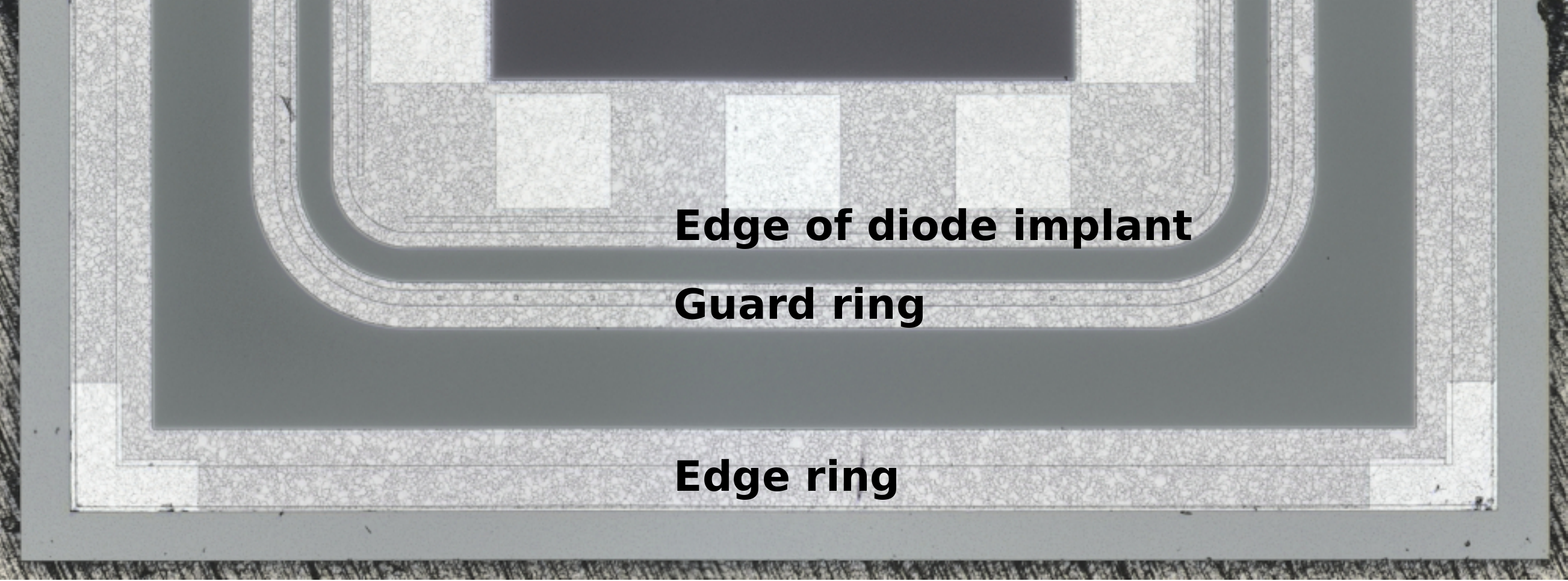}
  \caption{edge region of diode HPK MD}
  \label{fig:edge1}
\end{subfigure}
\begin{subfigure}{.67\textwidth}
  \centering
  \includegraphics[width=\linewidth]{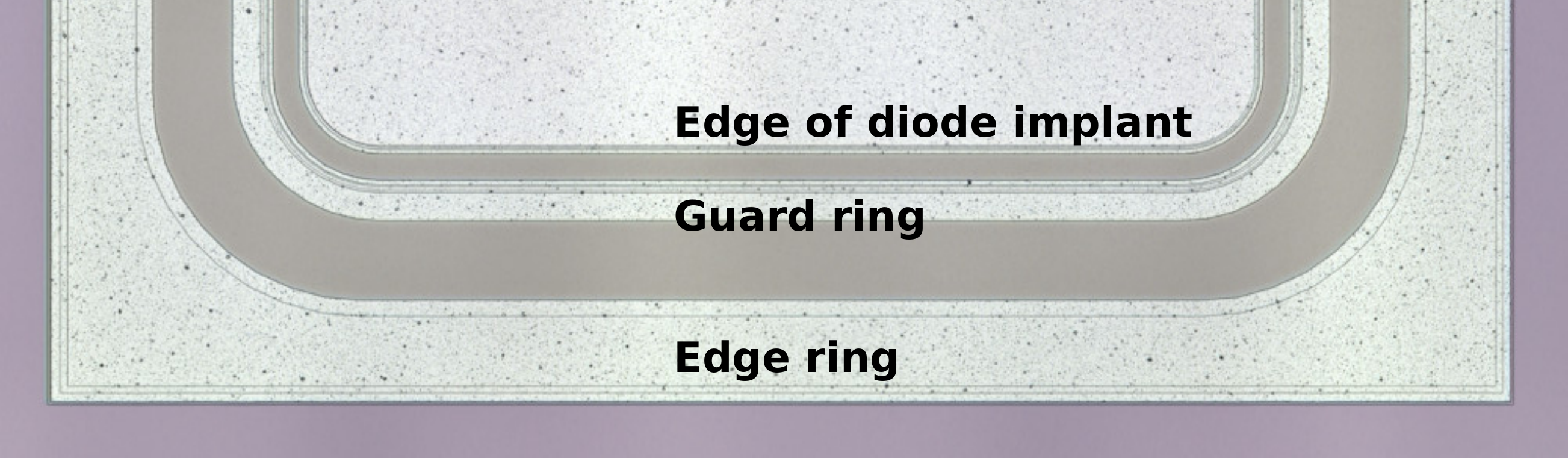}
  \caption{edge region of diode IFX MD2}
  \label{fig:edge2}
\end{subfigure}
 \begin{subfigure}{\textwidth}
  \centering
  \includegraphics[width=\linewidth]{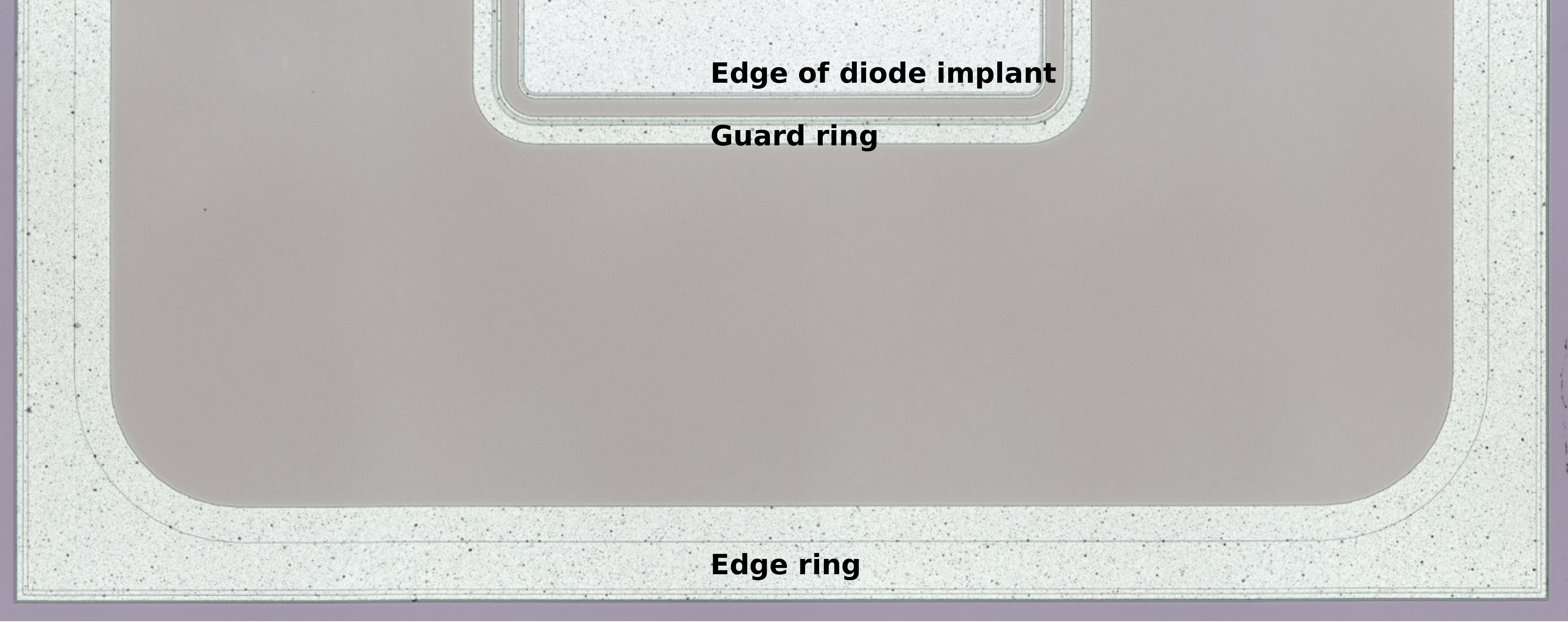}
  \caption{edge region of diode IFX TD3}
  \label{fig:edge3}
\end{subfigure}
\caption{Edge regions of diodes under investigation: diode implants and guard rings of diodes HPK MD and IFX MD2 can be seen to have similar shapes and sizes. Different from the straight angles of the HPK MD edge ring, IFX MD2 and TD3 diode edge rings show rounded corners.}
\label{fig:edges}
\end{figure}

Three diodes from two different wafers, produced by Hamamatsu Photonics K.K. (HPK) and Infineon Technologies (IFX), were studied (see figure~\ref{fig:test}). Diode HPK MD was on a wafer with a bulk resistivity of \unit[3.0]{k$\Omega\cdot$cm}, diodes IFX MD2 and IFX TD3 were on a wafer with a bulk resistivity of \unit[3.5]{k$\Omega\cdot$cm}. Depletion voltages for devices from the same wafers were measured to be about \unit[-300]{V}.
In addition to two $\unit[2\times2]{\text{mm}^2}$ diodes with a comparable layout of bias and guard rings (HPK MD and IFX MD2), a diode with a smaller active area, but larger distance to the dicing edge was chosen (IFX TD3) for comparison. While all diodes under investigation had a bias ring and guard ring similar to full size sensors, their edge rings vary in shape and position with respect to the active diode area (see figures~\ref{fig:edge1} to~\ref{fig:edge3}).

\section{Mapping the electric field using a micro-focused X-ray beam}

The aim of this study was to map the depleted area inside biased diodes with respect to the positions of field shaping design features (bias ring, guard ring, edge ring) and study its extension toward the dicing edges. Since the field shape in different depths of the diodes was of less interest than its extension in the sensor plane, this measurement was not set up to map the diode edge (e.g. by performing an edge-TCT measurement), but the field distribution with respect to the sensor plane: a micro-focused \unit[15]{keV} X-ray beam was arranged normal to the diode surface and the response at different diode positions measured.

Each \unit[15]{keV} photon has a \unit[51]{\%} chance to interact within silicon with a thickness of \unit[300]{$\upmu$m}. An interaction produces one \unit[15]{keV} electron, which travels up to \unit[20]{$\upmu$m} in silicon and produces about 4,200 electron-hole-pairs. 

The Diamond Light Source provides photons in intervals of \unit[2]{ns}, which the process of producing a micro-focused, monochromatic beam reduces to a photon flux of about \unit[$4.7\pm0.8$]{photons} per \unit[10]{ns}. Given the integration time used for current measurements (\unit[20]{ms}, see section~\ref{sec:scans}), variations on the time scale of \unit[10]{ns} can be neglected.

While in the absence of an electric field, electrons and holes recombine, the presenceof an electric field leads to the free charge carriers moving towards the sensor surface and back plane. They thereby cause an increase in the measured leakage current due to induced ionisation produced by electrons from interactions with X-ray photons.

It should be mentioned that due to the unknown interaction depth of each X-ray photon, the current measured for each diode position is integrated over the full sensor depth. This measurement therefore does not provide information about the field distribution in different diode depths, as e.g. a Two-Photon-Absorption (TPA)-TCT measurement would. 

Each diode under investigation was moved with respect to the beam using precision translation stages. By measuring the resulting current for each position, the electric field inside the diode was mapped.
In these measurements, all diodes were mounted on circuit boards designed for beam tests of test structures. High voltage was applied to the diode back plane using silver conductive paint (LS 200 N) and its surface was grounded using a wire bond (see figures~\ref{fig:dioden_1} to~\ref{fig:dioden_3}). 

Guard rings were floating during the performed measurements as is foreseen for the operation of sensors in the ATLAS ITk. The impact of a floating guard ring was previously studied in probe station measurements of a full-size ATLAS12 sensor, which was operated with both a floating guard ring (only the bias ring was connected to ground) and grounded guard ring (bias ring and guard ring were connected to the same ground). No difference was found in the sensor leakage current for measurements with a grounded or floating guard ring.

Each diode was cooled to \unit[0]{$^{\circ}$C} and kept at below \unit[1]{\%} humidity in a light-tight box during the measurements.

For these measurements, a monochromatic \unit[15]{keV} X-ray beam, micro-focused to a size (FWHM) of $\unit[1.8\times 3.2]{\upmu\text{m}^2}$ using a compound refractive lens, was provided by beam line B16 at the Diamond Light Source~\cite{B16}.

\section{Performed scans}
\label{sec:scans}

\begin{figure}
\centering
\includegraphics[width=0.7\linewidth]{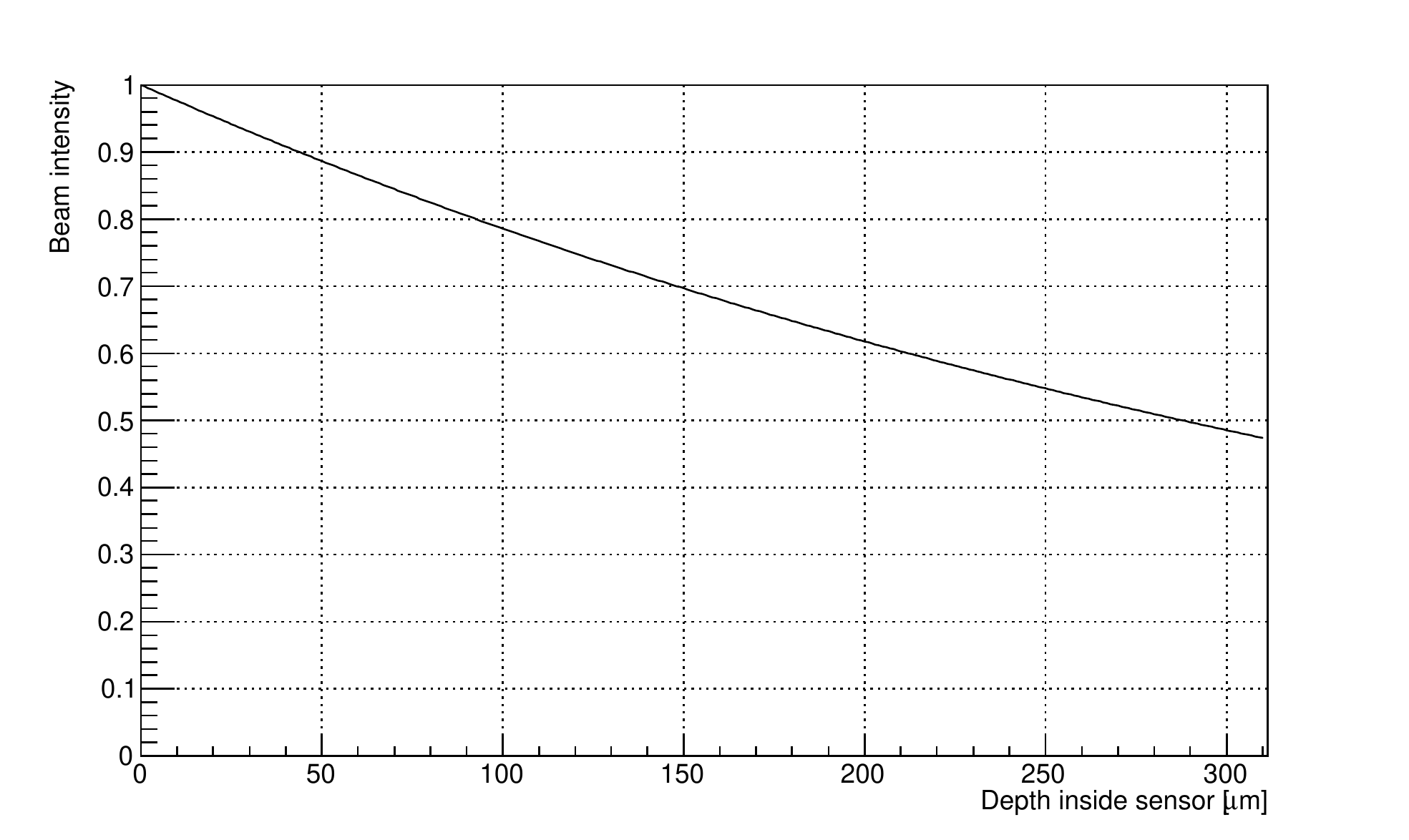}
\caption{Intensity of a {\unit[15]{keV}} electron beam while traversing a silicon sensor with a thickness of {\unit[310]{$\upmu$m}}.}
\label{fig:intensity}
\end{figure}

For each diode, three scans were performed:
\begin{enumerate}
 \item a coarse scan in large steps over the full diode area in order to map its outer edges
 \item a fine scan in small steps over one or two corners of each diode
 \item line scans through the centre of each diode in both horizontal and vertical direction for different bias voltages to map the dependence of the electric field on the applied bias voltage.
\end{enumerate}
Line scans were set up to use large step sizes over the plateau area of the diode centre and small steps in the more interesting edge regions. Voltages were varied between \unit[-50]{V} (corresponding to a less than half depleted sensor) and \unit[-500]{V} (corresponding to an over-depleted sensor) to study the diode response for different depletion depths. 
It should be mentioned that due to absorption, the number of photons traversing the diode decreases exponentially. Since the diode was positioned facing the sensor, more photon interactions occur in the upper region of the diode than in the lower regions, where the photon beam intensity has decreased (see figure~\ref{fig:intensity}).
Since the sensor depletes from the surface downwards, a larger fraction of photons interacts in the upper, depleted diode volume than in the lower, un-depleted volume, which results in a non-liner current increase with increasing depletion depth.
In the following, scans parallel to the dicing edges of a diode will be referred to as "across" (horizontal/vertical) and scans through opposite corners of a diode as "diagonal".
Table~\ref{tab:scans} summarises the positioning parameters used for the scans of each diode.
\begin{table}[htp]
\centering
\begin{tabular}{l|l|c|c|c}
 & & \multicolumn{3}{c}{Diode under investigation} \\
\hline
\rotatebox[origin=c]{90}{\parbox[c]{1cm}{\centering Scan}} & & HPK MD & IFX MD2 & IFX TD3 \\
\hline
 \multirow{5}{*}{\rotatebox[origin=c]{90}{\parbox[c]{1cm}{\centering Coarse}}} & Length in $x$, $[$mm$]$ & 2.1 & 2.6 & 3.4 \\
 & Step size in $x$, $[$mm$]$ & 0.1 & 0.1 & 0.15 \\
 & Length in $y$, $[$mm$]$ & 2.1 & 2.4 & 3.1 \\
 & Step size in $y$, $[$mm$]$ & 0.1 & 0.1 & 0.15 \\
 & Bias voltage, $[$V$]$ & -400 & -400 & -400 \\
 \hline
 \multirow{7}{*}{\rotatebox[origin=c]{90}{\parbox[c]{1cm}{\centering Fine}}} & Number of scans & 2 & 1 & 2 \\
 & Remarks & 2 corners & large area & attenuation\\
 & Length in $x$, $[$mm$]$ & 0.55 & 2.4 & 1.0 \\
 & Step size in $x$, $[\upmu$m$]$ & 20 & 25 & 50 \\
 & Length in $y$, $[$mm$]$ & 0.55 & 1.2 & 1.0 \\
 & Step size in $y$, $[\upmu$m$]$ & 20 & 25 & 50 \\
 & Bias voltage, $[$V$]$ & -400 & -400 & -400 \\
 \hline
 \multirow{8}{*}{\rotatebox[origin=c]{90}{\parbox[c]{1cm}{\centering Lines}}} & Length in $x$, $[$mm$]$ & 2.1 & 2.4 & 2.8 \\
 & Step size in $x$ (fine), $[\upmu$m$]$ & 10 & 10 & 50 \\
 & Step size in $x$ (coarse), $[\upmu$m$]$ & 100 & 100 & - \\
 & Length in $y$, $[$mm$]$ & 2.2 & 2.4 & 2.9 \\
 & Step size in $y$ (fine), $[\upmu$m$]$ & 10 & 10 & 50 \\
 & Step size in $y$ (coarse), $[\upmu$m$]$ & 100 & 100 & - \\
 & Min$/$max voltage, $[$V$]$ & -50/500 & -50/-500 & -50/-500 \\
 & Bias voltage steps, $[$V$]$ & 50 & 50 & 50 \\
 \end{tabular}
\caption{Parameters for scans performed for three diodes under investigation. For IFX TD3, fine scans of the same diode corner were performed using two different beam intensities by inserting an attenuator in the photon beam.}
\label{tab:scans}
\end{table}
At each position, the current was measured using the power supply readout of a Keithley 2410 high voltage power supply, which was also used to bias the diode under investigation. A python script was used to read the current from the power supply after a waiting time of \unit[3]{s} following a stage movement, from an average calculated from 20 measurements with an integration time of \unit[20]{ms} each.

\section{Results}

All results show the absolute current measured at each position and therefore include induced photo current as well as the dark current. The dark current of each diode was measured to be $\unit[\leq1]{\text{nA}}$ for bias voltages up to \unit[-500]{V}.

\subsection{Coarse scans}

\begin{figure}
\centering
\includegraphics[width=\linewidth]{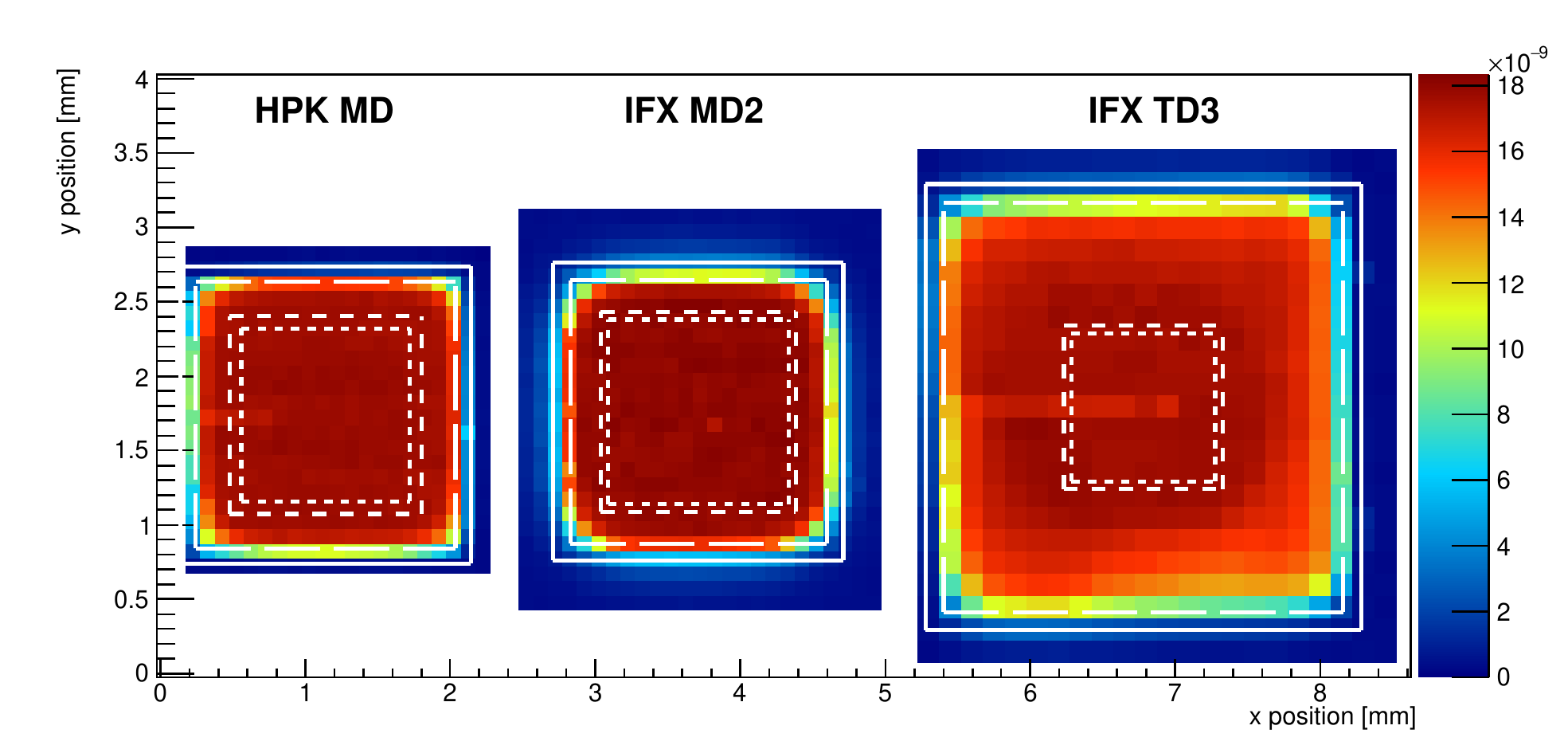}
\caption{Results of coarse scans (step sizes of {\unit[100]{$\upmu$m}} and {\unit[150]{$\upmu$m}}) of three diodes under investigation in comparison (to scale). White lines indicate approximate positions of features of each diode: outline of diode implant (dotted line), guard ring (small dashed line), centre of edge ring (large dashed line) and nominal size of each diode (solid line, {\unit[$2\times2$]{mm$^2$}} and {\unit[$3\times3$]{mm$^2$}}). Areas with a lower measured current, which are located on the plateau at the diode centre, correspond to the positions of wire bonds on each diode, which reduce the beam intensity by about {\unit[5]{\%}} (corresponding to {\unit[25]{$\upmu$m}} of aluminium).}
\label{fig:coarse1}
\end{figure}

Figure~\ref{fig:coarse1} shows maps of all diodes under investigation based on coarse scans.
Currents measured at the centre of each diode were found to be similar, as can be expected from devices with similar characteristics and thicknesses. Plateaus at the centre of each diode were found to remain flat beyond the sensor bias ring towards the edges of the diode.
%The asymmetric shape of the current profile measured for IFX TD3 can be attributed to the position of the diode at the wafer edge, where the doping profile can vary in concentration and therefore affect the measured current distribution.

\begin{figure}
\centering
\begin{subfigure}{.49\textwidth}
  \centering
  \includegraphics[width=\linewidth]{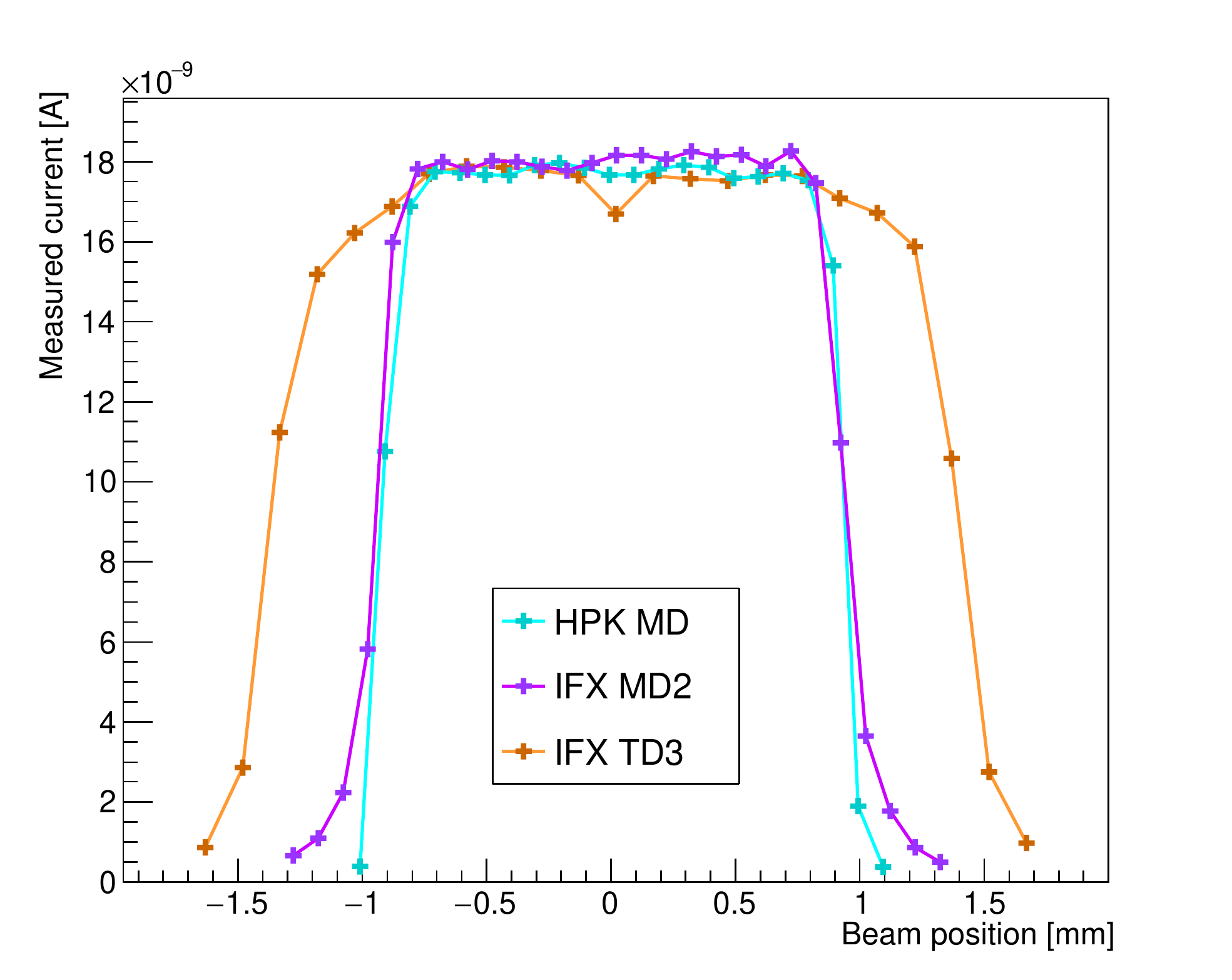}
  \caption{Parallel-scan current profiles (for vertical scan line) through the centre of each diode}
  \label{fig:coarse2}
\end{subfigure}
\begin{subfigure}{.49\textwidth}
  \centering
  \includegraphics[width=\linewidth]{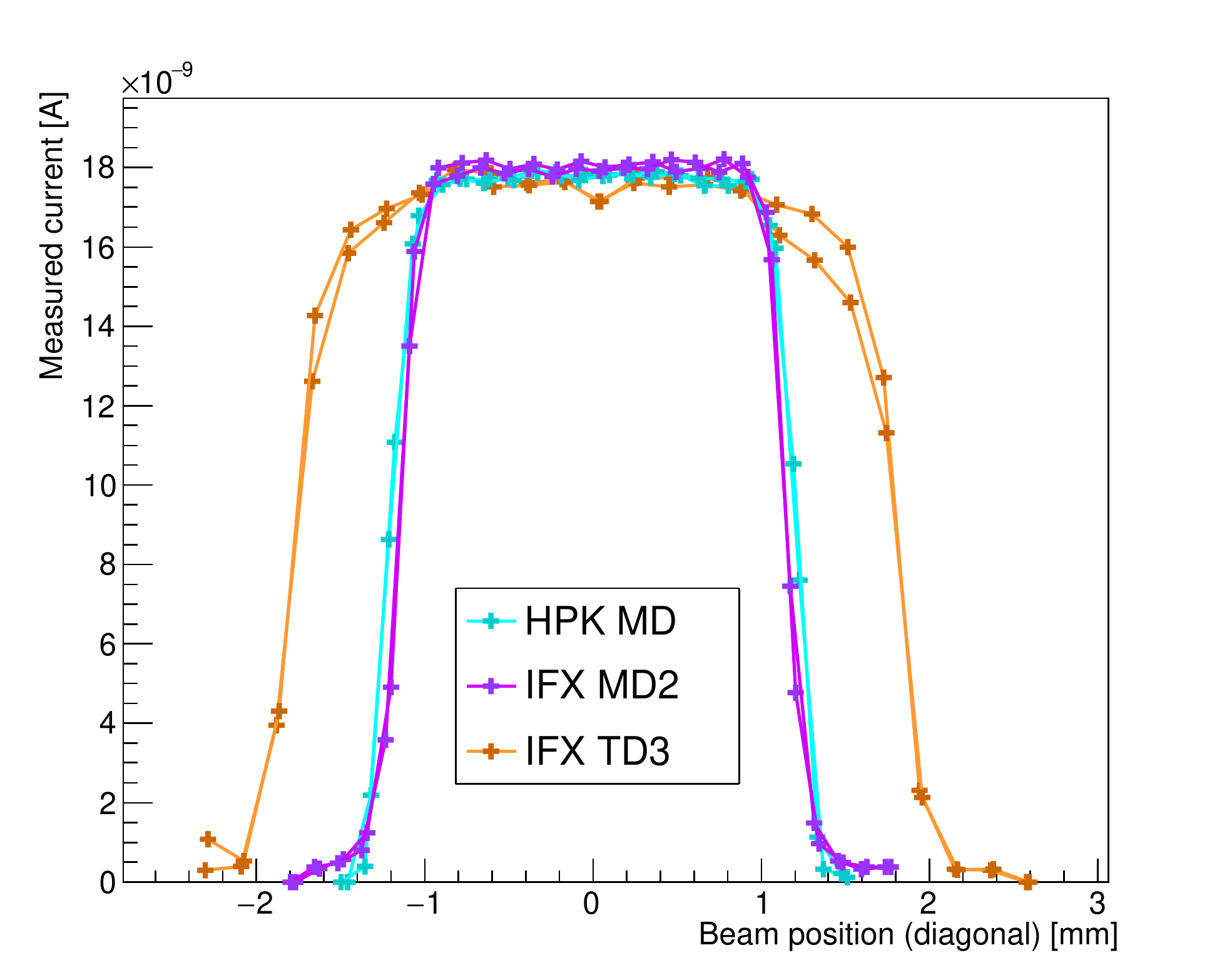}
  \caption{Diagonal-scan current profiles through the centre (for both pairs of opposite corners) of each diode}
  \label{fig:coarse3}
\end{subfigure}
\caption{Current profiles of slices through the centre of each diode. While HPK MD and IFX MD2 show similar profiles from edge to edge, profiles through diode corners show an extension further toward the diode corner for the HPK diode, matching the shape of its edge ring (see figure~\ref{fig:edge1}). A local minimum at the centre of diode IFX TD3 is caused by the presence of a wire bond, which reduces the beam intensity.}
\label{fig:coarse}
\end{figure}

Diodes HPK MD and IFX MD2 show similar current profiles, i.e. electric fields, towards the diode edges. They differ in the extension of the field towards the diode corners (see figures~\ref{fig:coarse2} and~\ref{fig:coarse3}), matching the different shapes of their edge rings (see figures~\ref{fig:edge1} and~\ref{fig:edge2}): the rectangular shaped edge ring of HPK MD leads to a wider extension of its electric field towards the diode corner than the rounded corners of the IFX MD2 edge ring. 

\begin{figure}
\centering
\includegraphics[width=0.9\linewidth]{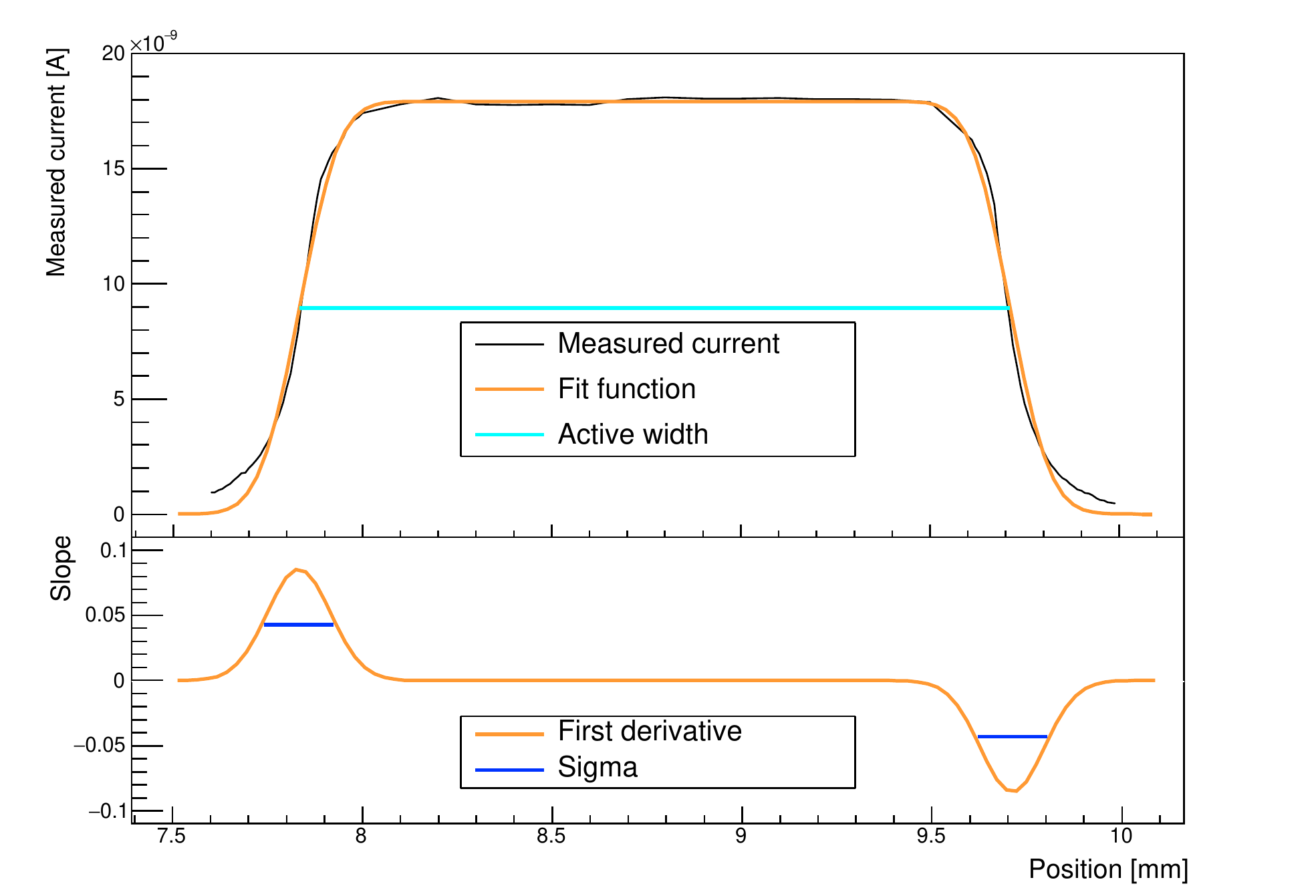}
\caption{Example of a beam current measurement (diode IFX MD2 at a bias voltage of {\unit[-300]{V}}) with applied fit function: the active width of each diode was calculated from the full width at half maximum, sigma was calculated from the slopes on each side of the plateau.}
\label{fig:sigma}
\end{figure}

The width and slope shape for each diode were calculated by applying a rectangular fit function $f_{\text{rect}}(x)$ to the measured current profile:
\begin{equation} 
f_{\text{rect}}(x) = \bigg[\bigg(-\frac{f_{\text{erf}}(-w/2 + a - x)}{\sqrt{2}\cdot \sigma}\bigg) + \bigg(\frac{f_{\text{erf}}(w/2 + a - x)}{\sqrt{2}\cdot \sigma}\bigg) \bigg] \cdot \frac{s}{2\cdot w},
\end{equation}
where $f_{\text{erf}}(x)$ is the gaussian error function, $w$ is the total width of the rectangular function, $a$ is the position of its centre, $s$ scales the height of the plateau and $\sigma$ is the width of the gaussian function.
Figure~\ref{fig:sigma} shows an example of a fit function applied to a diode measurement.

Table~\ref{tab:coarse} summarises the characteristics determined for each diode.
\begin{table}[htp]
\centering
\begin{tabular}{l|c|c|c}
 & \multicolumn{3}{c}{Diode under investigation} \\
 & HPK MD & IFX MD2 & IFX TD3 \\
 \hline
 Nominal width (across), $[$mm$]$ & 2 & 2 & 3 \\
 Active width (across), $[$mm$]$ & 1.85 & 1.96 & 2.63\\
 Percentage (measured/nominal) & \unit[92.5]{\%} & \unit[98.0]{\%} & \unit[87.7]{\%} \\
% Nominal width (bias ring), $[$mm$]$ & 1.14 & 1.25 & 1.00\\
% Extension (outside bias ring), $[$mm$]$ & 0.34 & 0.34 & 0.82 \\
 Sigma (across), $[$mm$]$ & 0.08 & 0.13 & 0.16 \\
 \hline
 Nominal width (diagonal), $[$mm$]$ & 2.83 & 2.83 & 4.24 \\
 Active width (diagonal), $[$mm$]$ & 2.37 & 2.32 & 3.55 \\
 Percentage (Measured/nominal) & \unit[83.7]{\%} & \unit[82.0]{\%} & \unit[83.7]{\%} \\ Sigma (diagonal), $[$mm$]$ & 0.08 & 0.11 & 0.18 \\
 \end{tabular}
\caption{Active width (measured) and sigma, i.e. width of gaussian function, from current profiles measured for diodes: diode HPK MD showed a narrower plateau across the diode with a smaller transition region than IFX MD2.}
\label{tab:coarse}
\end{table}
The measurements show that the electric field inside a depleted volume follows the shape of the edge ring around the active area.
%
%In addition to profiles, the obtained maps allow estimation of the size of the depleted diode area by integrating over all cells of the current plateau, i.e. with a high measured current. For reference, the average current at the centre of each diode was calculated and all bins for which a current close to the average current at the diode centre had been read out, were integrated (see figure~\ref{fig:areas}).
%\begin{figure}
%\centering
%\includegraphics[width=0.7\linewidth]{areas.eps}
%\caption{Depleted area of each diode under investigation as a fraction of the nominal diode area inside the bias ring, as a function of the required minimum current (relative to the average current measured at the centre of each diode). Uncertainties were calculated based on the assumption that {\unit[50]{\%}} of each cell on and around the outline of a constructed plateau could be wrongly attributed to the depleted or undepleted diode area.}
%\label{fig:areas}
%\end{figure}
%It was found that both MD2 diodes show different fractions of depleted areas with respect to the nominal area enclosed by the bias ring, with HPK showing a larger depleted area. For diode IFX TD3, the fraction of depleted diode area showed a larger dependence on the minimum required current used to attribute cells to either the depleted or undepleted diode area, as well as a significantly larger depleted area in general.

\subsection{Fine scans}

\begin{figure}
\centering
\includegraphics[width=\linewidth]{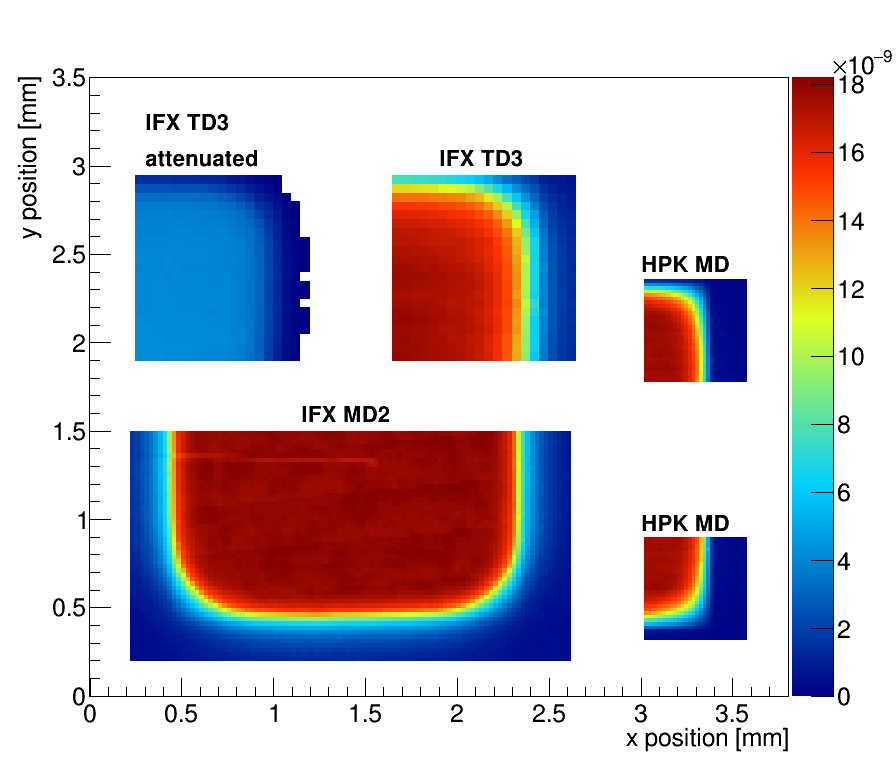}
\caption{Fine scans (step sizes of {\unit[20]{$\upmu$m}}, {\unit[25]{$\upmu$m}} and {\unit[50]{$\upmu$m}}) of corner regions of all diodes under investigation (to scale). The current profile shape matches the geometry of the edge ring (see figure~\ref{fig:test}): while the rounded corners of edge rings on IFX MD2 and IFX TD3 lead to a similarly rounded shape of the current profile in the diode corners, the rectangular shape of the edge ring on HPK MD leads to the shape of its current profile extending further towards the corners. The distance between both corners of HPK MD represents the actual distance between both measurement areas.}
\label{fig:fine}
\end{figure}

\begin{figure}
\centering
\includegraphics[width=0.7\linewidth]{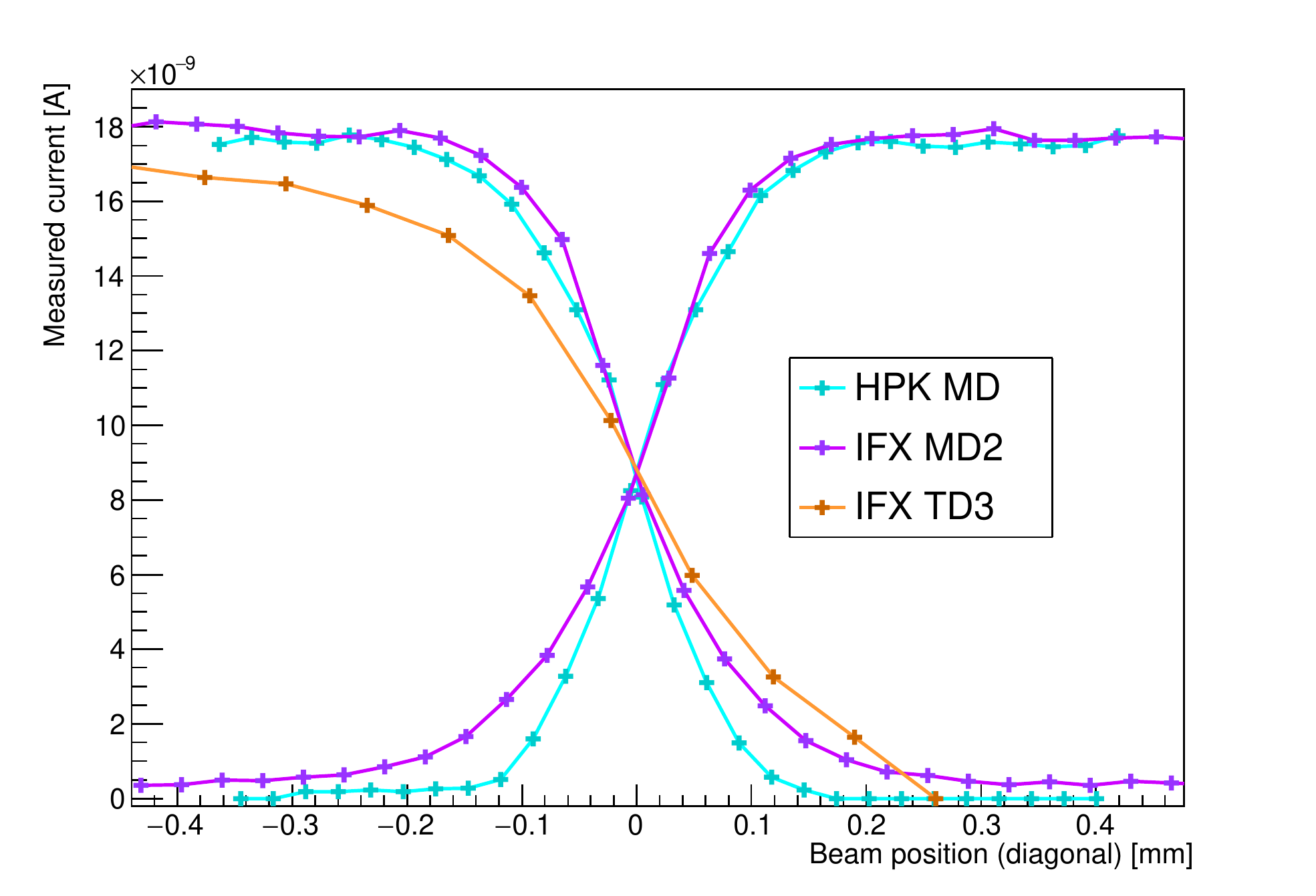}
\caption{Diagonal current profiles from slices through diode corners. Finely spaced scan points show a flatter slope of IFX TD3 compared to the other sensors and a steeper slope for HPK MD compared to IFX MD2. Profiles were moved to arrange the centre of each slope at $x = 0$ for better visual comparability.}
\label{fig:fine1}
\end{figure}

Fine scans of corner regions were performed to study several details observed in the previous scan. Figure~\ref{fig:fine} shows an overview of all fine-scan results for comparison.
As indicated in the coarse scans of diodes HPK MD and IFX MD2, the fine scan confirmed that the electric field of HPK MD extends further towards the diode corner than diode IFX MD2 while showing a steeper slope (see figure~\ref{fig:fine1}).

\begin{figure}
\centering
\begin{subfigure}{.90\textwidth}
  \centering
  \includegraphics[width=\linewidth]{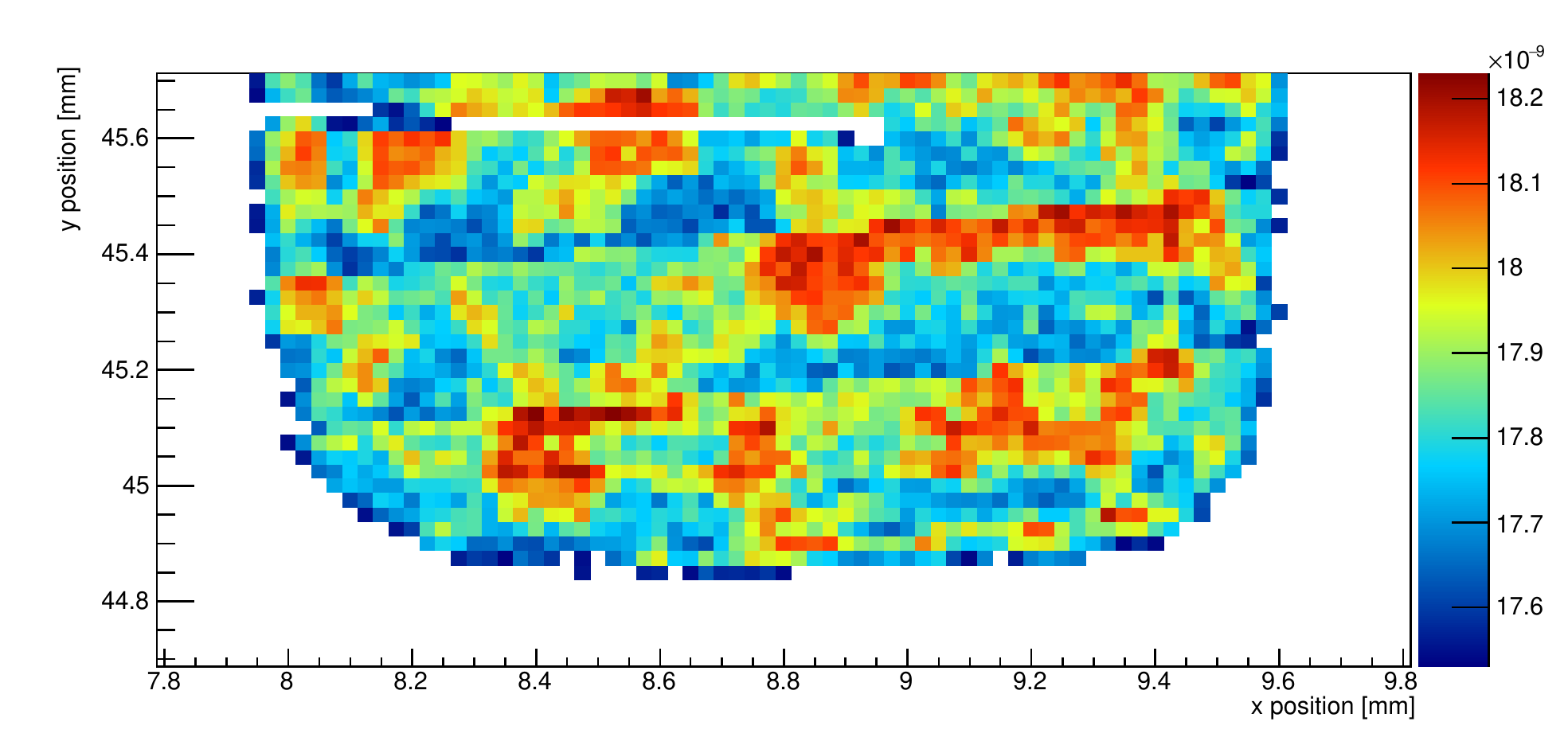}
  \caption{Plateau region of diode IFX TD3, scanned in steps of {\unit[25]{$\upmu$m}}. The measured current shows fluctuations of up to {\unit[0.3]{nA}}.}
  \label{fig:var1}
\end{subfigure}
\begin{subfigure}{.9\textwidth}
  \centering
  \includegraphics[width=\linewidth]{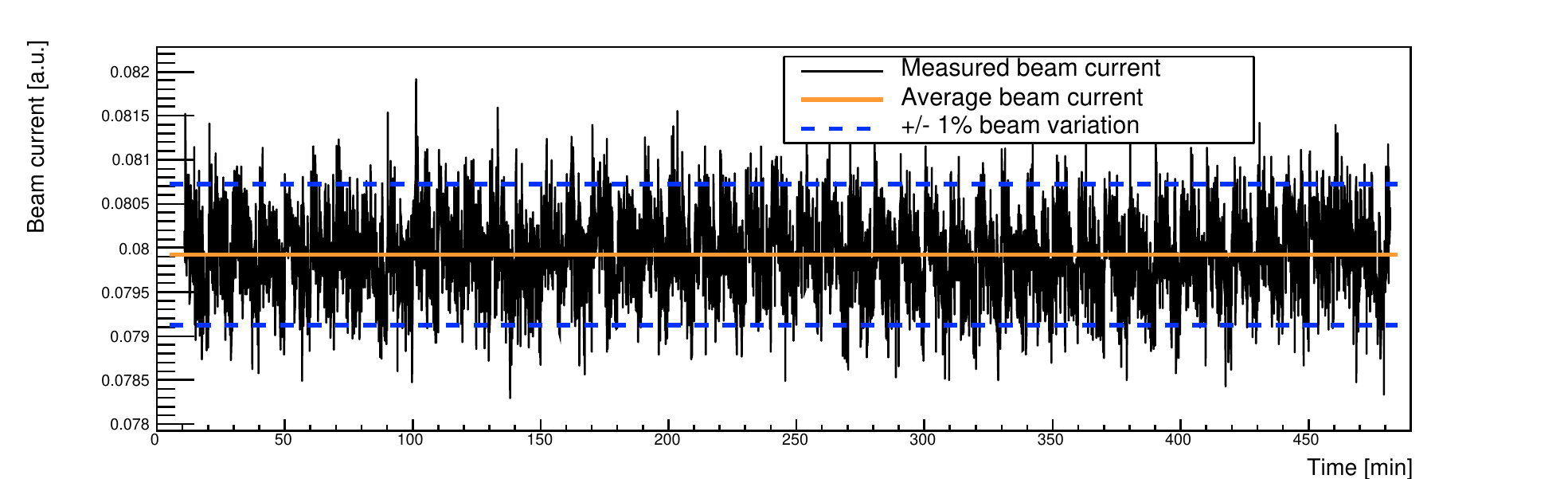}
  \caption{Measured beam current over time: measurements show fluctuations of more than {\unit[$\pm1$]{\%}}.}
  \label{fig:var2}
\end{subfigure}
\caption{Fluctuations in the current at the central region of diode IFX TD3 and the corresponding beam current measured during the scan. Both measurements show similar variations.}
\label{fig:var}
\end{figure}

Current measured at different positions on the plateau of each diode showed variations of \unit[0.3]{nA} throughout the plateau (see figure~\ref{fig:var1}), which corresponds to \unit[$\pm1.7$]{\%} of the total current.
Monitoring the beam current during the test showed variations on a similar scale to that of the currents measured for the diode. Current variations throughout the plateau of a diode can therefore be attributed to beam current variations.

\subsection{Line scans}
\label{subsec:lines}

\begin{figure}
\centering
\begin{subfigure}{.45\textwidth}
  \centering
  \includegraphics[width=\linewidth]{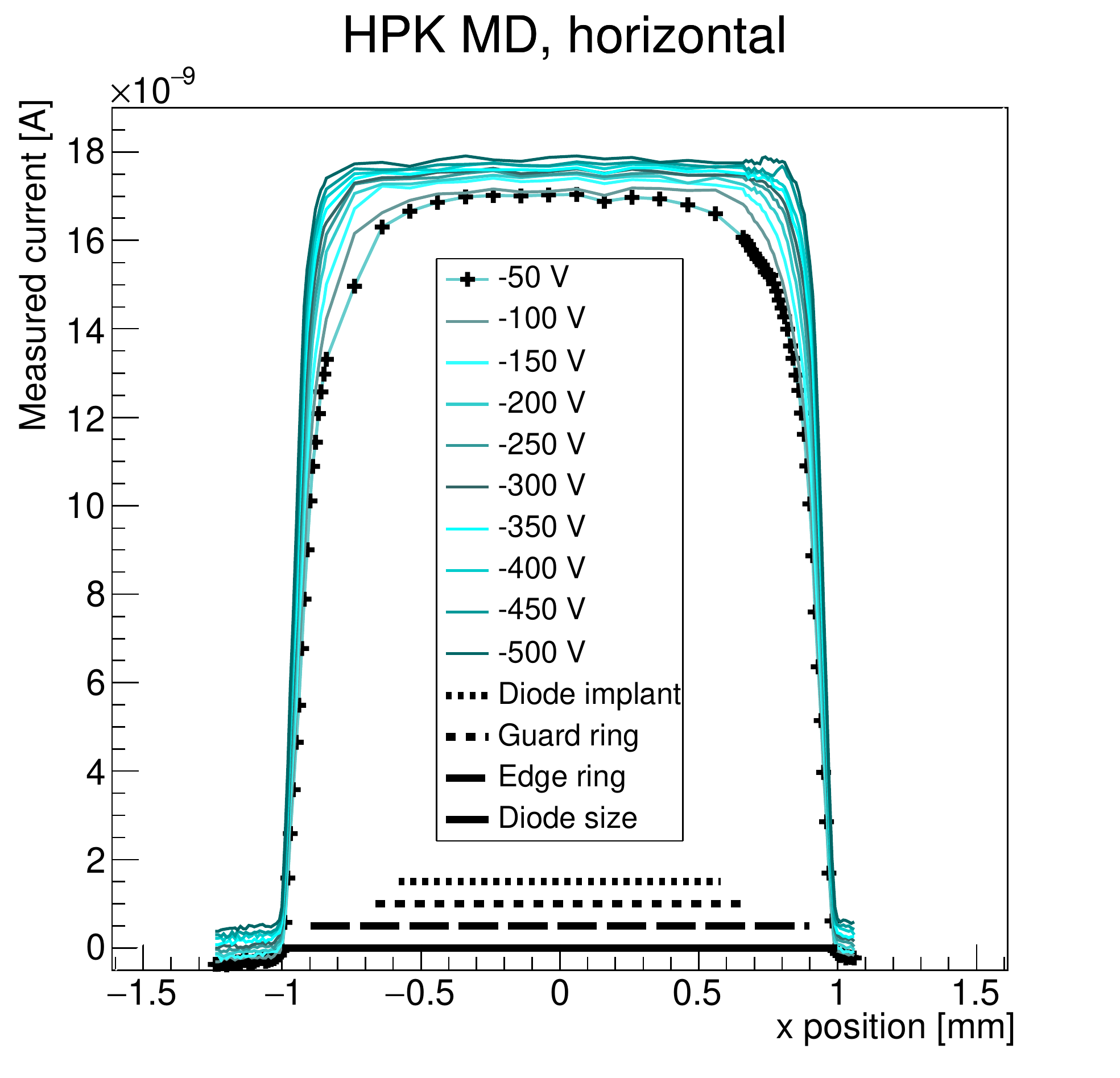}
  \caption{HPK MD, horizontal}
  \label{fig:cross2}
\end{subfigure}
\begin{subfigure}{.45\textwidth}
  \centering
  \includegraphics[width=\linewidth]{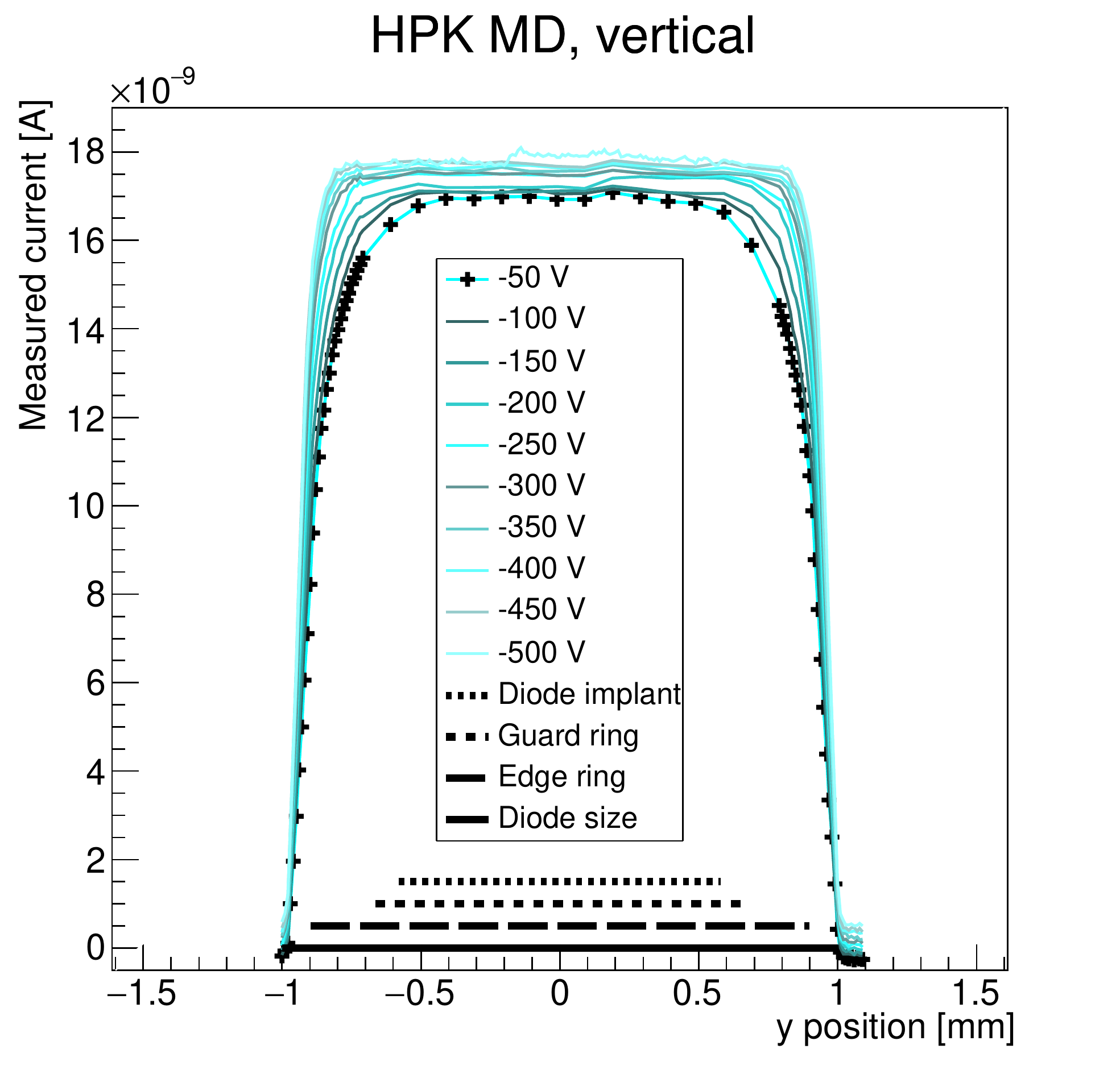}
  \caption{HPK MD, vertical}
  \label{fig:cross3}
\end{subfigure}
\begin{subfigure}{.45\textwidth}
  \centering
  \includegraphics[width=\linewidth]{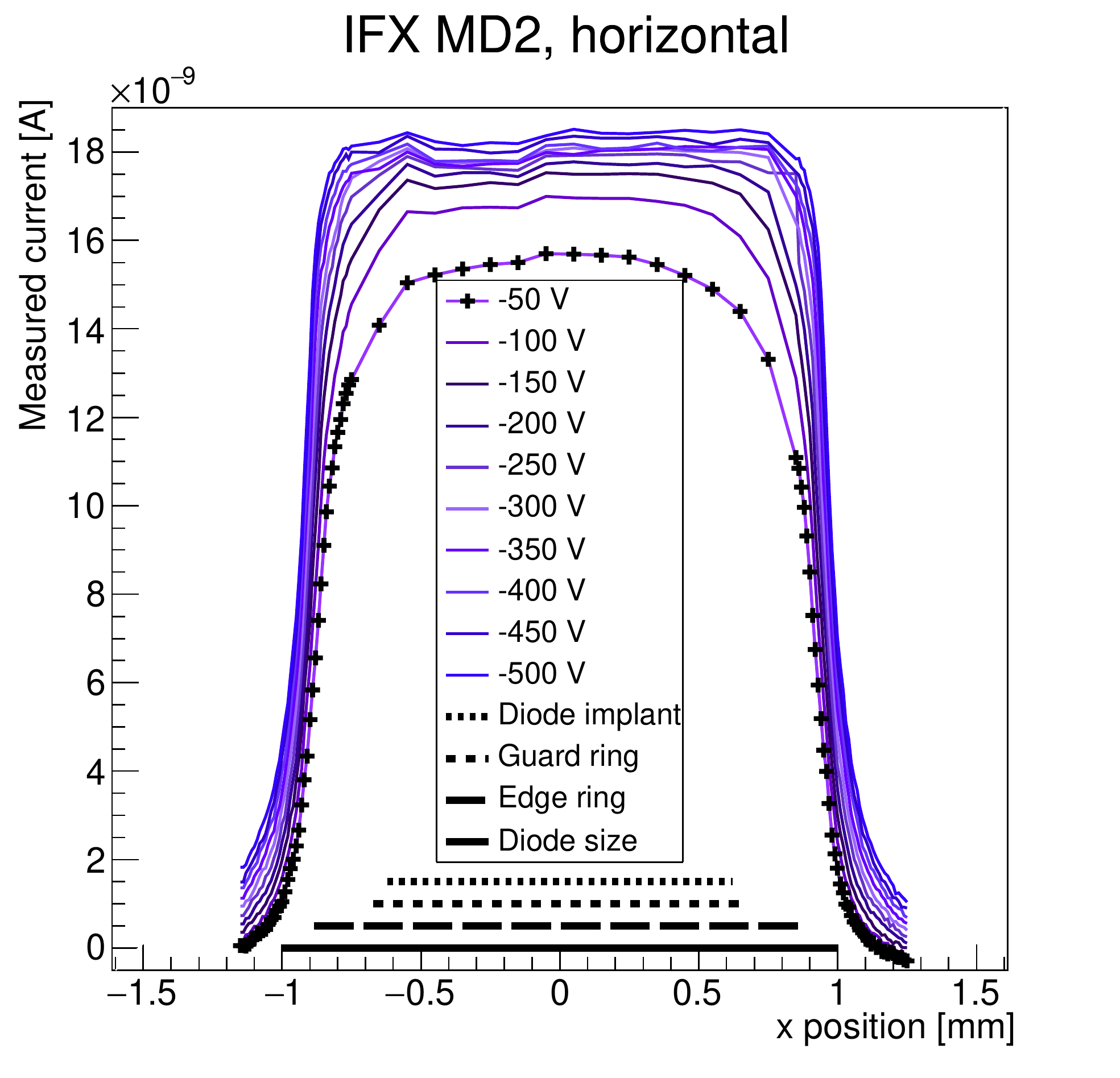}
  \caption{IFX MD2, horizontal}
  \label{fig:cross4}
\end{subfigure}
\begin{subfigure}{.45\textwidth}
  \centering
  \includegraphics[width=\linewidth]{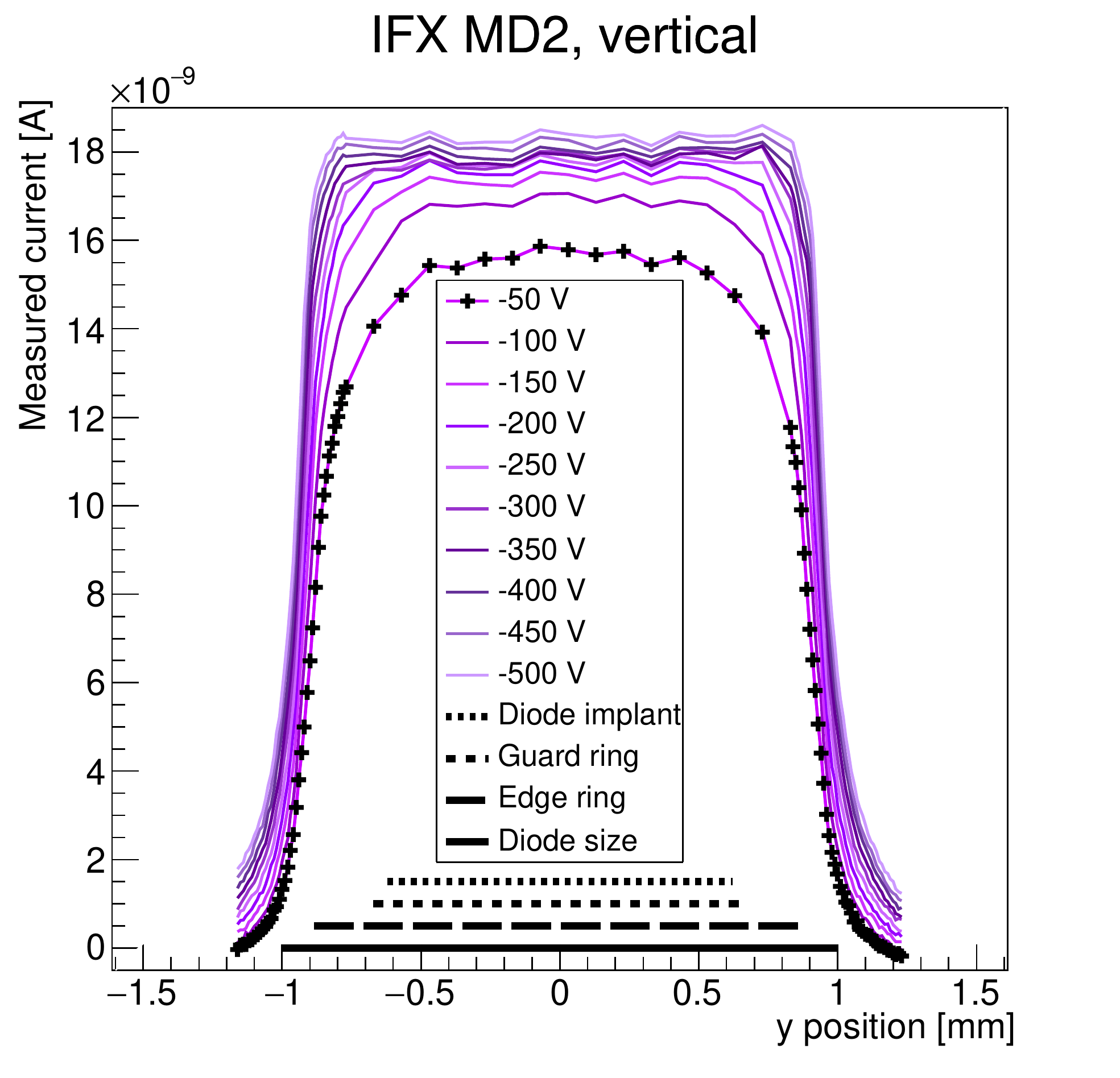}
  \caption{IFX MD2, vertical}
  \label{fig:cross5}
\end{subfigure}
\begin{subfigure}{.45\textwidth}
  \centering
  \includegraphics[width=\linewidth]{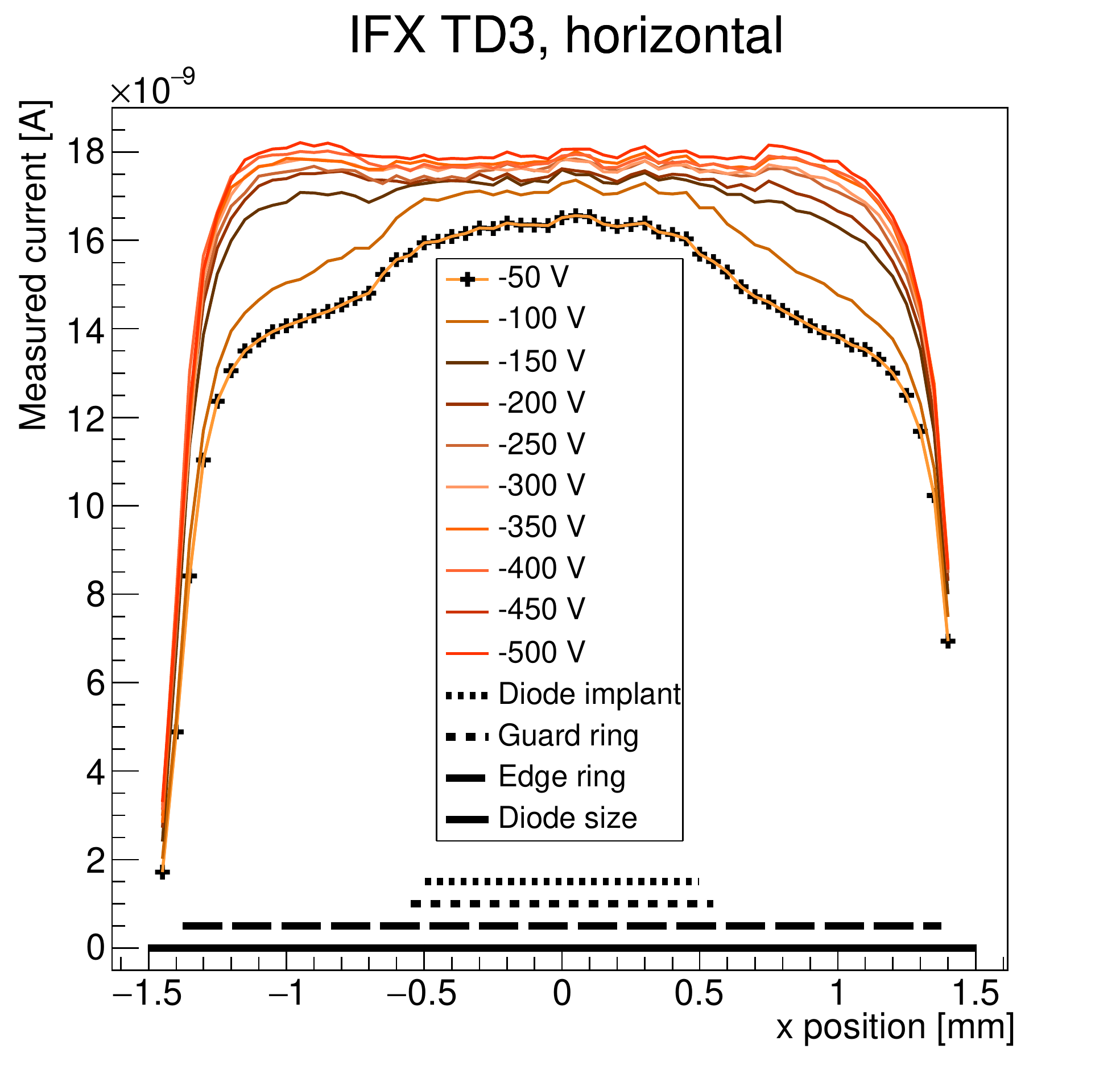}
  \caption{IFX TD3, horizontal}
  \label{fig:cross6}
\end{subfigure}
\begin{subfigure}{.45\textwidth}
  \centering
  \includegraphics[width=\linewidth]{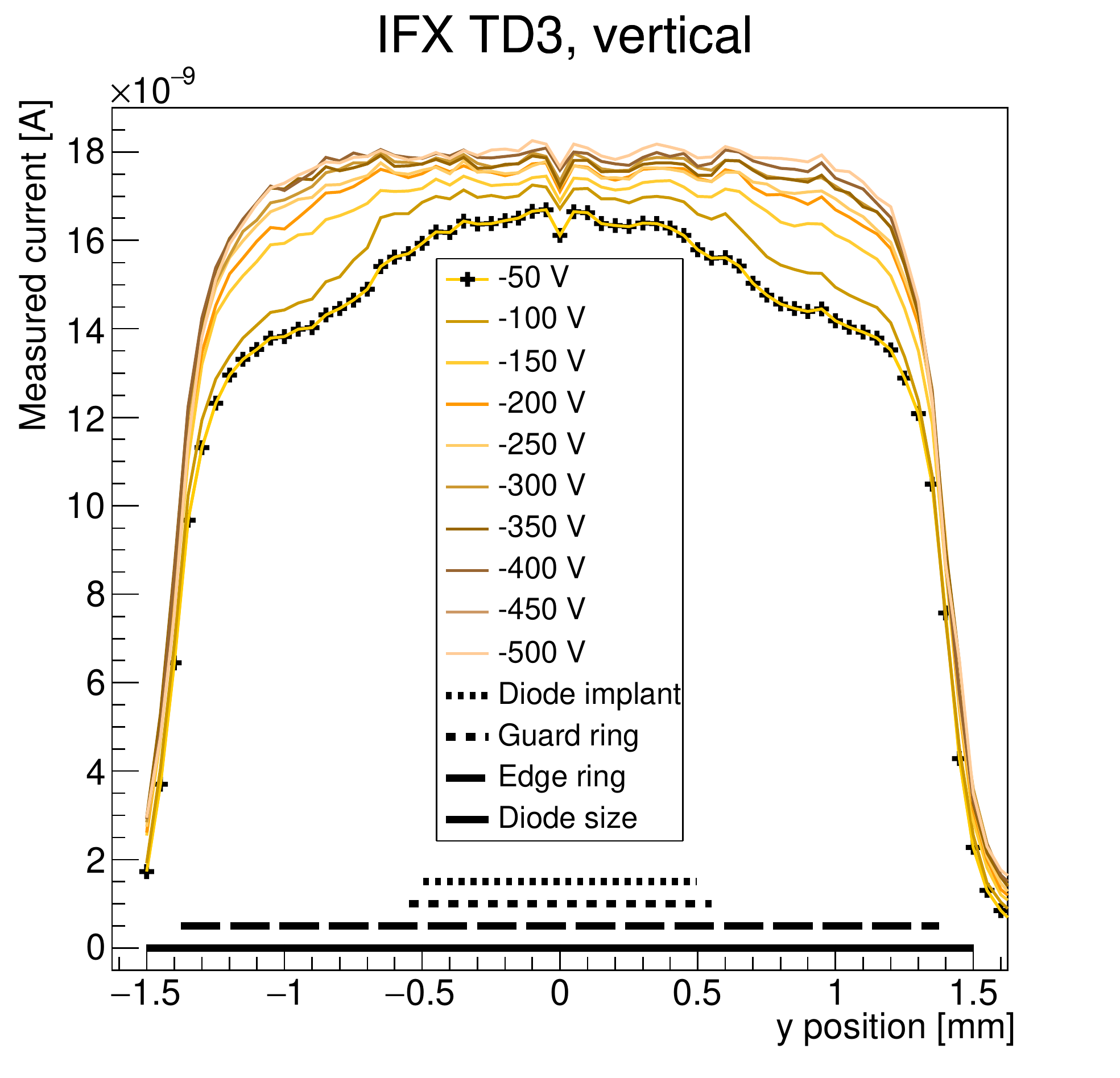}
  \caption{IFX TD3, vertical}
  \label{fig:cross7}
\end{subfigure}
\caption{Results from line scans across diodes at different bias voltages. Markers indicate the positions of individual scan points. Black lines indicate diameters of diode features of each diode: diode implant (dotted line), guard ring (small dashed line), centre of edge ring (large dashed line), nominal diode size (solid line).}
\label{fig:cross}
\end{figure}

Line scans over the centre of each diode at different bias voltages confirmed that the width of the depleted sensor area is mostly independent of the applied bias voltage (see figures~\ref{fig:cross2} to~\ref{fig:cross7}). For all diodes, increasing bias voltages led to the measured current showing a flatter plateau, i.e. an extension of the electric field further towards the diode edges.

It is interesting to note that although the nominal depletion voltage for sensors from this wafer is about \unit[-300]{V}, measured current profiles show only minor changes above voltages of \unit[-200]{V}. This effect is assumed to be caused by the decreasing beam intensity between the diode surface and back plane (see figure~\ref{fig:intensity}).
HPK MD and IFX MD2 diodes registered different currents for low bias voltages, with diode IFX MD2 showing a significantly lower read out current at \unit[-50]{V}.

Only diode IFX TD3 showed a dependence of the read out current on the position of the guard ring: for low bias voltages (\unit[-50]{V} and \unit[-100]{V}), the diode shows a plateau corresponding to the size of the diode implant. For higher bias voltages, the plateau extends over most of the diode width up to the edge ring.
%The asymmetry of line scans across the IFX TD3 diode is again assumed to be the result of the doping profile changing towards the wafer edge.
Line scans confirmed the extension of the electric field is defined by the diode edge ring.

\section{Conclusion and Outlook}

Scans in different areas of three diodes under investigation and at different bias voltages confirmed that the electric field inside silicon with high bulk resistivity is shaped by the diode edge ring. The width of the depleted area is almost independent of the applied bias voltage (beyond \unit[-200]{V}).

The depleted area inside the diodes under investigation was found to reach the edge ring, where its depth was reduced over a slope length from \unit[80]{$\upmu$m} (HPK MD) to \unit[130]{$\upmu$m} (IFX TD3) (see figure~\ref{fig:sigma}). The results confirm that an edge ring with sufficient width and doping concentration has enough field stopping power to prevent a contact between the electric field inside a semiconductor and its highly degraded dicing surface.

Future measurements are planned to be conducted on irradiated devices in order to compare the extension of the electric field before and after irradiation.

\section*{Acknowledgements}

We thank the Diamond Light Source for access to beam line B16 (proposal number MT18807) that contributed to the results presented here. The authors would like to thank the personnel of beam line B16, especially Oliver Fox, for providing advice, support and maintenance during the experiment.

Individual authors$^1$ were supported in part by the U.S. Department of Energy under Contract No. DE-AC02-05CH11231.
The work at SCIPP$^4$ was supported by the Department of Energy, grant DE-SC0010107.
This work$^5$ is supported and financed in part by the Spanish Ministry of Science, Innovation and Universities through the Particle Physics National Program, ref. FPA2015-65652-C4-4-R (MICINN/FEDER, UE), and co-financed with FEDER funds.

\bibliographystyle{unsrt}
\bibliography{bibliography.bib}

\begin{thebibliography}{1}

\bibitem{ATLAS}
{The~ATLAS~Collaboration}.
\newblock {The ATLAS Experiment at the CERN Large Hadron Collider}.
\newblock {\em Journal of Instrumentation}, 3(08):S08003, 2008.

\bibitem{ITk}
{The~ATLAS~Collaboration}.
\newblock {Technical Design Report for the ATLAS Inner Tracker Strip Detector}.
\newblock Technical Report CERN-LHCC-2017-005. ATLAS-TDR-025, CERN, Geneva, Apr
  2017.

\bibitem{ATLAS07}
Y.~Unno et~al.
\newblock Development of n-on-p silicon sensors for very high radiation
  environments.
\newblock {\em {Nuclear Instruments and Methods in Physics Research Section A:
  Accelerators, Spectrometers, Detectors and Associated Equipment}}, 636(1,
  Supplement):S24 -- S30, 2011.
\newblock {7th International Hiroshima Symposium on the Development and
  Application of Semiconductor Tracking Detectors}.

\bibitem{ATLAS12}
L.~B.~A. Hommels et~al.
\newblock {Detailed studies of full-size ATLAS12 sensors}.
\newblock {\em Nucl. Instrum. Meth.}, A831:167--173, 2016.

\bibitem{ENEXSS}
{Semiconductor Leading Edge Technologies, Inc. (SELETE)}.
\newblock {\em {Hyper Environment for Exploration of Semiconductor Simulation
  (ENEXSS)}}, 2006/4–2011/3.
\newblock {developed by a consortium}.

\bibitem{B16}
K.~J.~S. Sawhney et~al.
\newblock {A Test Beamline on Diamond Light Source}.
\newblock {\em {AIP Conference Proceedings}}, 1234(1):387--390, 2010.

\end{thebibliography}

\end{document}